\documentclass[preprints,communication,accept,moreauthors,pdftex]{mdpi} 

\firstpage{1} 
\pubvolume{xx}
\issuenum{1}
\articlenumber{5}
\pubyear{2019}
\copyrightyear{2019}
\history{Received: 30 April 2019; Accepted: 23 May 2019; Published: date}

\pdfoutput=1

\setitemize{parsep=6pt,itemsep=0pt,leftmargin=*,labelsep=5.5mm}
\setenumerate{parsep=6pt,itemsep=0pt,leftmargin=*,labelsep=5.5mm}
\setlist[description]{itemsep=0mm}

\usepackage{soul}

\Title{Multiplicity Dependence in the Non-Extensive Hadronization Model Calculated by the HIJING++~Framework}

\Author{G\'abor B\'ir\'o $^{1,2,*,\dagger}$\orcidA{}, Gergely G\'abor Barnaf\"oldi $^{1,*,\dagger}$\orcidC{}, G\'abor Papp $^{2,*,\dagger}$\orcidB{} and Tam\'as S\'andor Bir\'o $^{1,*,\dagger}$\orcidD{}
}

\AuthorNames{G\'abor B\'ir\'o, Gergely G\'abor Barnaf\"oldi, G\'abor Papp, Tam\'as S\'andor Bir\'o}

\address{%
$^{1}$ \quad Wigner Research Centre for Physics of the Hungarian Academy of Sciences, Department of Theory, 29-33 Konkoly-Thege Mikl\'os Str, H-1121 Budapest, Hungary\\
$^{2}$ \quad Institute for Physics, E\"otv\"os Lor\'and University, 1/A P\'azm\'any P. S\'et\'any, H-1117, Budapest, Hungary
}

\corres{{Correspondence: biro.gabor@wigner.mta.hu (G.B.); barnafoldi.gergely@wigner.mta.hu (G.G.B); pg@ludens.elte.hu (G.P.); biro.tamas@wigner.mta.hu (T.S.B.)
} }

\firstnote{These authors contributed equally to this work.}

\abstract{The non-extensive statistical description of the identified final state particles measured in high energy collisions is well-known by its wide range of applicability. However, there are many open questions that need to be answered, including but not limited to, the question of the observed mass scaling of massive hadrons or the size and multiplicity dependence of the model parameters. This latter is especially relevant, since currently the amount of   available experimental data with high multiplicity at small systems is very limited. This contribution has two main goals: On  the one hand we provide a status report of the ongoing tuning of the soon-to-be-released \texttt{HIJING++} Monte Carlo event generator. On the other hand, the role of multiplicity dependence of the parameters in the non-extensive hadronization model is investigated with \texttt{HIJING++} calculations. We present cross-check comparisons of \texttt{HIJING++} with existing experimental data to verify its validity in our range of interest as well as calculations at high-multiplicity regions where we have insufficient experimental data.
}
    
\keyword{high energy physics; heavy-ion; Monte Carlo; event generator; parallel computing; HIJING, non-extensive; Tsallis}

\begin{document}


\section{Introduction}
\label{sec:intro}

The transverse momentum {($p_T$)} distribution of identified hadrons stemming from high-energy proton--proton, proton--nucleus, and nucleus--nucleus collisions is one of the most fundamental observables in high-energy physics. In recent years, the Tsallis--Pareto-like distributions, motivated~from non-extensive statistical physics, have  received close attention because their applicability in this field~\cite{BIROG:TSALLIS2, GRIGORYAN:TSALLIS, ZHU:TSALLIS, CLEYMANS:TSALLIS, WONG:TSALLIS, TRIPATHY:TSALLIS, SHEN:TSALLIS2}. With the appearance of high precision experimental data spanning from low- to high-$p_T$, neither the thermal models with a bare Boltzmann--Gibbs exponential distribution nor {perturbative Quantum Chromodynamics (pQCD)}-motivated power-law distributions are able to describe the whole spectrum. On the other hand, the Tsallis--Pareto distributions combine these two regions perfectly (see, e.g., \cite{BIROG:TSALLIS1, BIROG:TSALLIS2, BIROG:TSALLIS3, BIROG:TSALLIS4, TAKACS:TSALLIS, GRIGORYAN:TSALLIS, URMOSSY:TSALLIS, ZHU:TSALLIS, VAN:TSALLIS, KHUNTIA:TSALLIS, WILK:TSALLIS, CLEYMANS:TSALLIS, SHEN:TSALLIS,WONG:TSALLIS, TRIPATHY:TSALLIS, SHEN:TSALLIS2} and references therein). During the investigation of the parameters, we showed that they possess non-trivial relations such as mass- and energy scaling~\cite{BIROG:TSALLIS1, BIROG:TSALLIS2}. There are also implications that for larger systems a \textit{soft-hard} extension is needed~\cite{BIROG:TSALLIS4, SHEN:TSALLIS}. These studies indicate that increasing the size of the colliding system (roughly speaking, the volume of the quark-gluon plasma) may also reflect  in the parameters. Our goal is therefore to systematically explore the parameter space as the function of the event multiplicity.

The \texttt{HIJING++} framework is a soon-to-be-published general purpose Monte Carlo event generator, currently in the final phase of development~\cite{HIJING, HIJING2, HPP1, HPP2, HPP3, HPP4}. It will serve as the successor of the \texttt{FORTRAN HIJING}, completely rewritten in modern C++. With the flexibility gained by using modular C++ structures, \texttt{HIJING++} also utilizes several external packages~\cite{PYTHIA, LHAPDF, GSL, VEGAS, ROOT, PDF1, PDF2}. Currently the internal parameters of \texttt{HIJING++} are being tuned to main experimental observables using \texttt{Professor}~\cite{PROFESSOR, PARALLEL}. This provides an excellent opportunity to test the capabilities of \texttt{HIJING++} and calculate high-multiplicity events.

In the next section, we briefly summarize the progress of tuning in \texttt{HIJING++} and present the current status. In Section \ref{sec:nonext}, a theoretical description of the transverse momentum spectra is given, and~the \texttt{HIJING++} calculations are given in Section \ref{sec:multiplicity}.
 
\section{Tuning of \texttt{HIJING++} Parameters}
\label{sec:tuning}

A typical general purpose Monte Carlo event generator, e.g., 
 the \texttt{HIJING++} framework, developed to be able to simulate high-energy heavy-ion collisions, has parameters that are not determined by theory and need to be tuned to reproduce measured experimental data with the highest possible precision. One of the main features of the \texttt{HIJING++} framework is that very few input parameters are needed to fully define a run, such as the species of the projectile and target beam and center-of-mass energy. Given this information, all of the other intrinsic parameters are calculated automatically.

Since \texttt{HIJING++} is based on the convolution of sequential collisions of nucleon--nucleon pairs in each nucleus--nucleus interaction, it is highly important to have a solid proton--proton {collisions} baseline. In this section, we present the up-to-date result of the tuning process using the following $\sqrt{s}=7$ TeV proton--proton experimental data:

\begin{itemize}
    \item $p_T$ spectra of identified $\pi^\pm$, $K^\pm$, and $p(\bar{p})$ hadrons with $INEL>0$ normalization (at least one charged particle in the $|\eta|<1.0$ region is required) up to $p_T=20$ GeV/c~\cite{ALICE:PID};
    \item charged hadron multiplicity distribution in the range of $\left<dN_{ch}/d\eta\right>=0-70$, {where $N_{ch}$ is the number of charged particles}~\cite{ALICE:MULTPICITY1, ALICE:MULTPICITY2};
    \item charged hadron {$\eta=\frac{1}{2}\ln{\frac{p+p_Z}{p-p_Z}}$} pseudorapidity distribution at {mid-pseudorapidity} $|\eta|<1.0$~\cite{ALICE:MULTPICITY1}.
\end{itemize}

The tuning process is performed iteratively utilizing the \texttt{Professor} tool~\cite{PROFESSOR, PARALLEL}. In Table \ref{tab:tune_parameters}, we list the main tunable parameters.
\begin{table}[H]
    \caption{Main internal parameters in \texttt{HIJING++}.}
    \label{tab:tune_parameters}
    \centering
    \tablesize{\footnotesize} 
    \begin{tabular}{cp{11.2cm}}
        \toprule
        \textbf{Parameter} & \textbf{Description} \\
        \midrule
    $p_0$               & soft-hard separation scale: minimum $p_T$ transfer of hard or semihard scatterings \\
    $\sigma_{soft}$ & the inclusive cross section for soft interactions \\
    $\sigma_0$  & the cross section that characterizes the geometrical size of a nucleon \\
   $\mu_0$  & the parameter in the scaled eikonal function of nucleon used to calculate total cross-section \\
   $K$  & K-factor for the differential jet cross sections in the lowest order pQCD calculation \\
\midrule
   $\max{p_T}_{cut}$ & $p_T$ cut for classifying the connected-independent type strings at fragmentation \\ 
   $m_{inv-cut}$ & invariant mass cut-off for the dipole radiation of a string system below which soft gluon radiations are terminated \\ 
   $m_{min-inv-ex.str.}$ & minimum value for the invariant mass of the excited string system in a hadron--hadron interaction \\ 
   $S_{p_{T_1}}$ & the parameter that regularizes the singularity at $p_T=0$ in the distribution of the soft $p_T$ kick \\ 
   $S_{p_{T_2}}$ & the parameter that gives the scale beyond which the $p_T$ kick distribution will be similar to $1/p_T^4$ \\ 
   $F$ & the scale in the form factor to suppress the $p_T$ transfer to diquarks in hard scatterings \\ 
\midrule
   $v_{q_i}$ & {phenomenological} parameters {($i=1,2,3$) of the soft parton distribution function} that yield an $x$ distribution of the valence quarks in a soft interaction \\
   $s_{q_i}$ & {phenomenological} parameters {($i=1,2,3$) of the soft parton distribution function} that yield an $x$ distribution of the sea quarks in a soft interaction \\
\midrule
   \texttt{StringPT:temperature} & the temperature parameter in the Lund fragmentation model as described in~\cite{PYTHIA}\\
   \texttt{StringPT:tempPreFactor} & the temperature prefactor for strange quarks and diquarks in the Lund fragmentation model as described in~\cite{PYTHIA}\\
   \texttt{StringZ:aExtraSQuark} & \multirow{2}{*}{parameters in the Lund symmetric fragmentation function as described in~\cite{PYTHIA}}\\
   \texttt{StringZ:aExtraDiuark} & \\ 
   \bottomrule
\end{tabular}
\end{table}

\textls[-15]{The detailed process of tuning and the parameters values will be described in the technical release paper of \texttt{HIJING++}. Here we present only the tuning status regarding the experimental data listed above.}

\begin{figure}[H]
    \centering
    \includegraphics[width=0.44\textwidth]{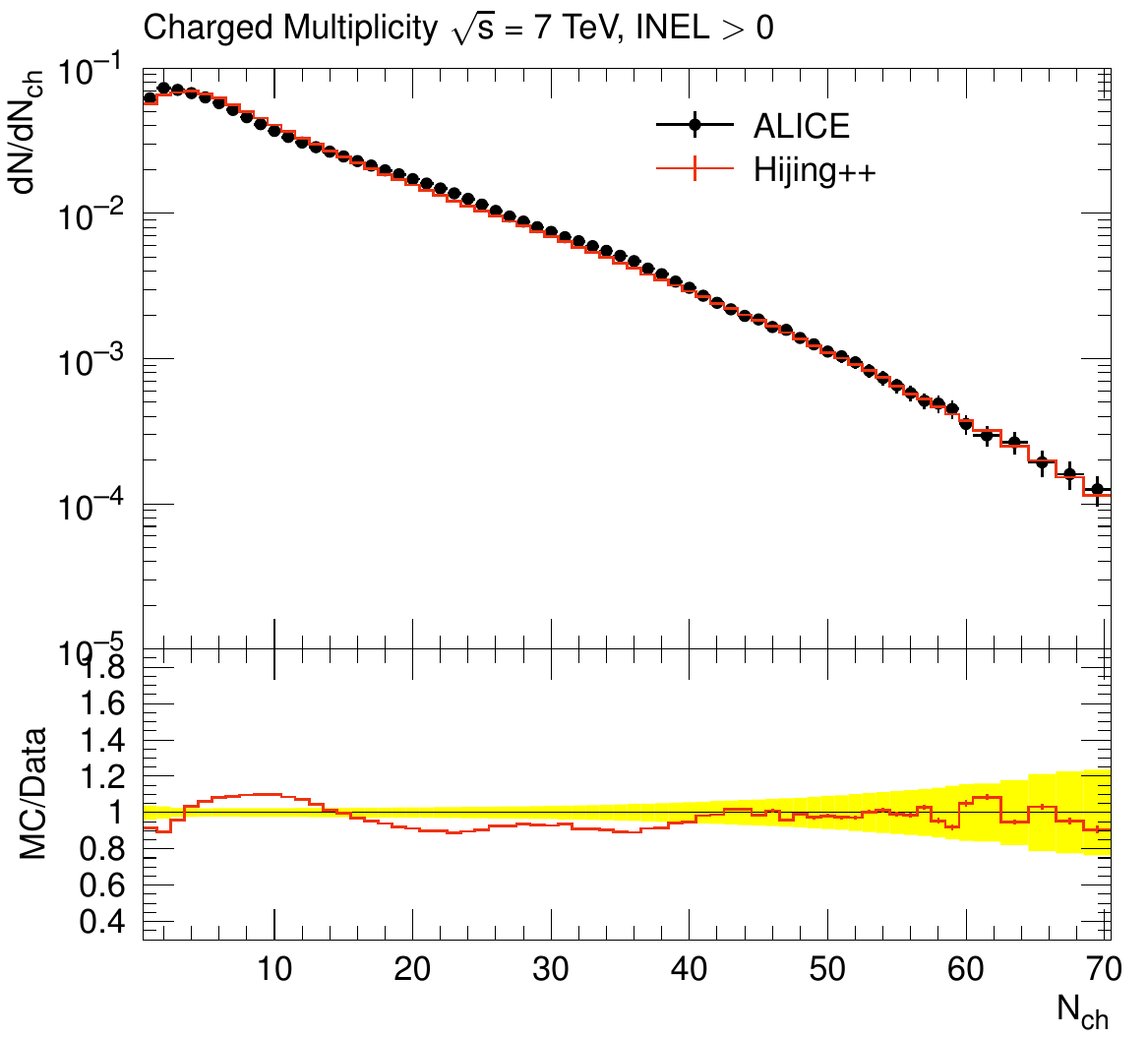}
    \includegraphics[width=0.44\textwidth]{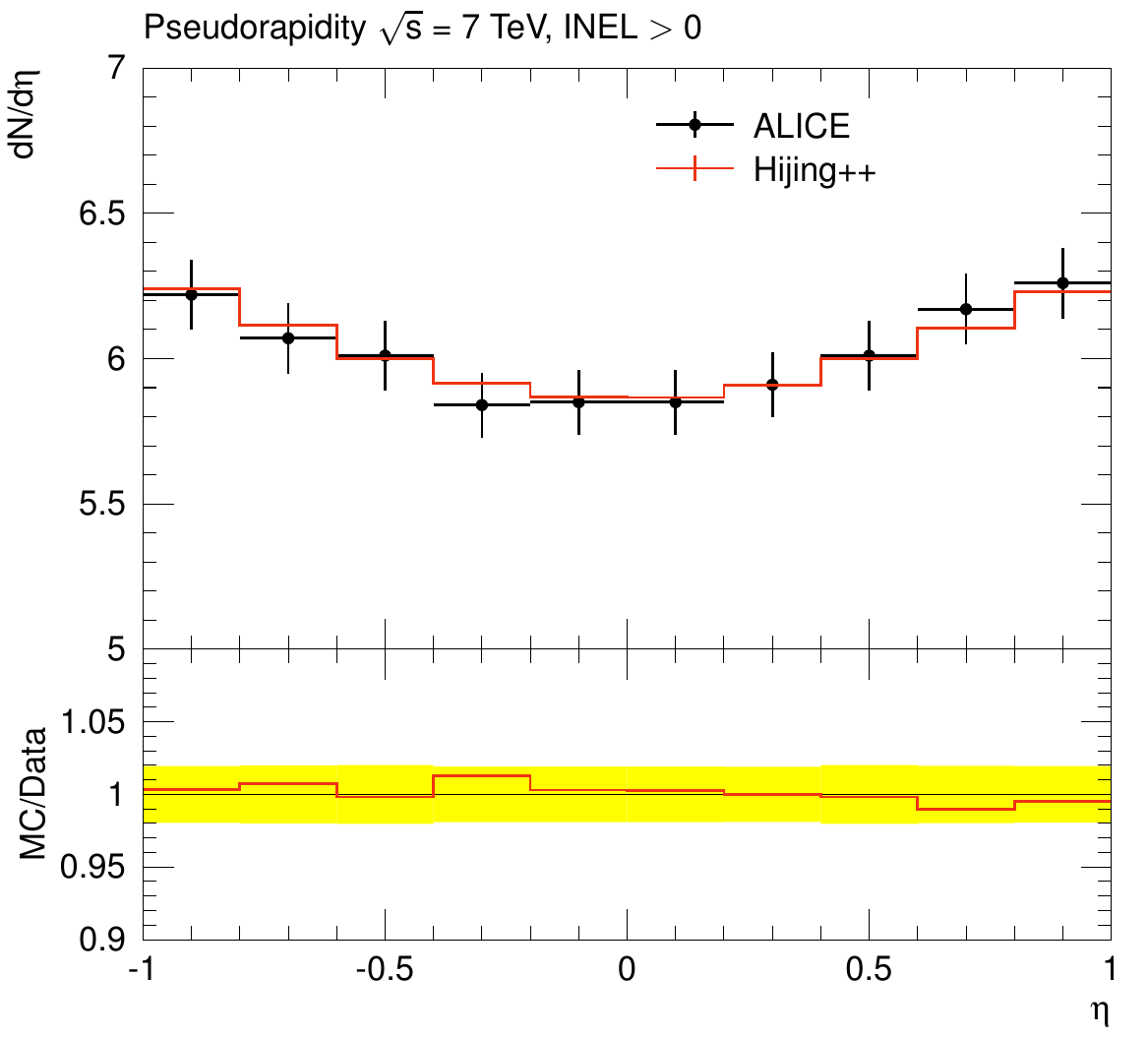}
    \caption{The multiplicity distribution (\textbf{{left panel}}) and pseudorapidity distribution (\textbf{{right panel
    }}) of charged hadrons stemming from proton--proton collisions at $\sqrt{s}=7$ TeV calculated with \texttt{HIJING++} and compared to experimental data~\cite{ALICE:MULTPICITY1, ALICE:MULTPICITY2}.}
    \label{fig:multiplicity}
  \end{figure}

\begin{figure}[H]
    \centering
    \includegraphics[width=0.32\textwidth]{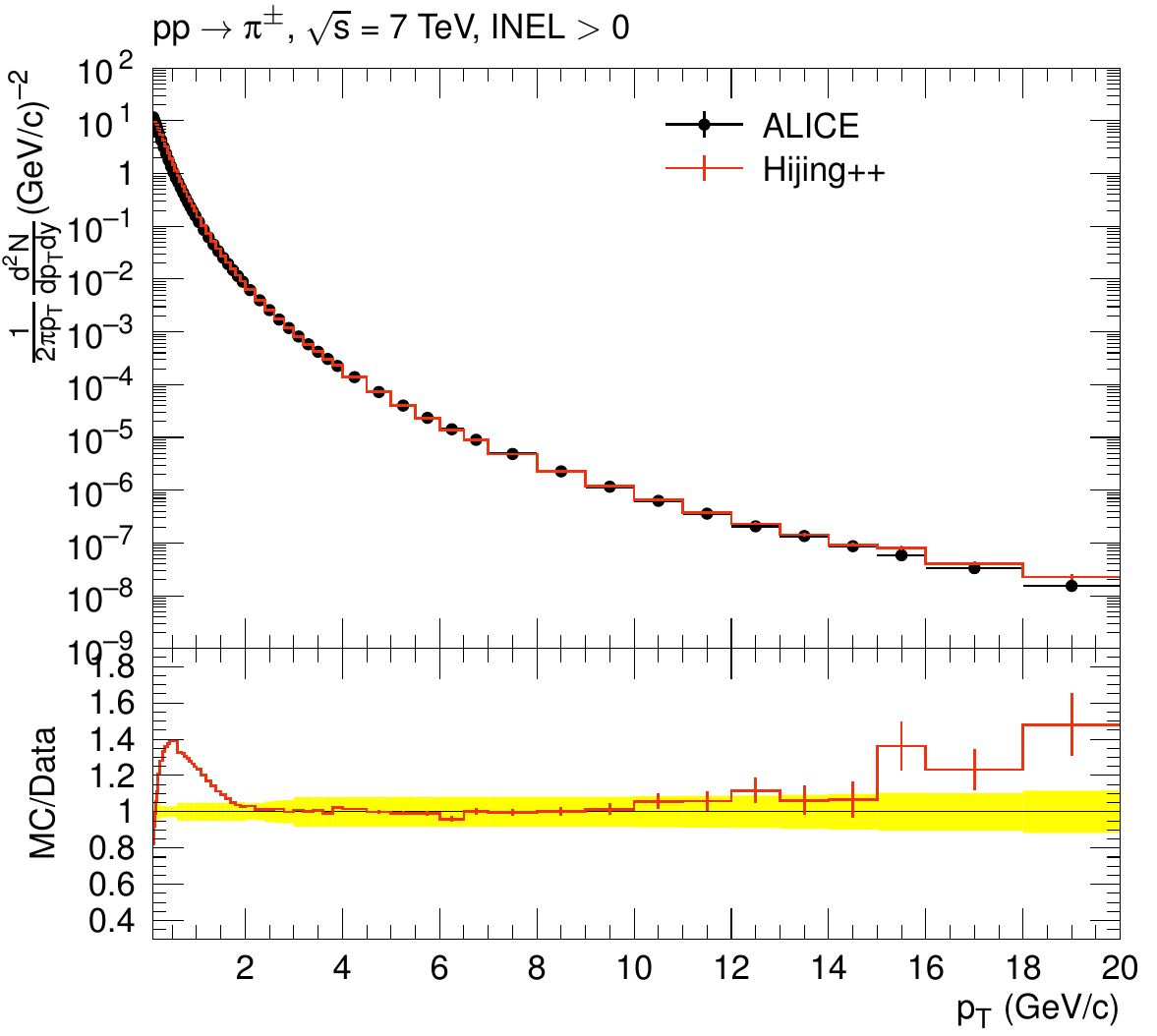}
    \includegraphics[width=0.32\textwidth]{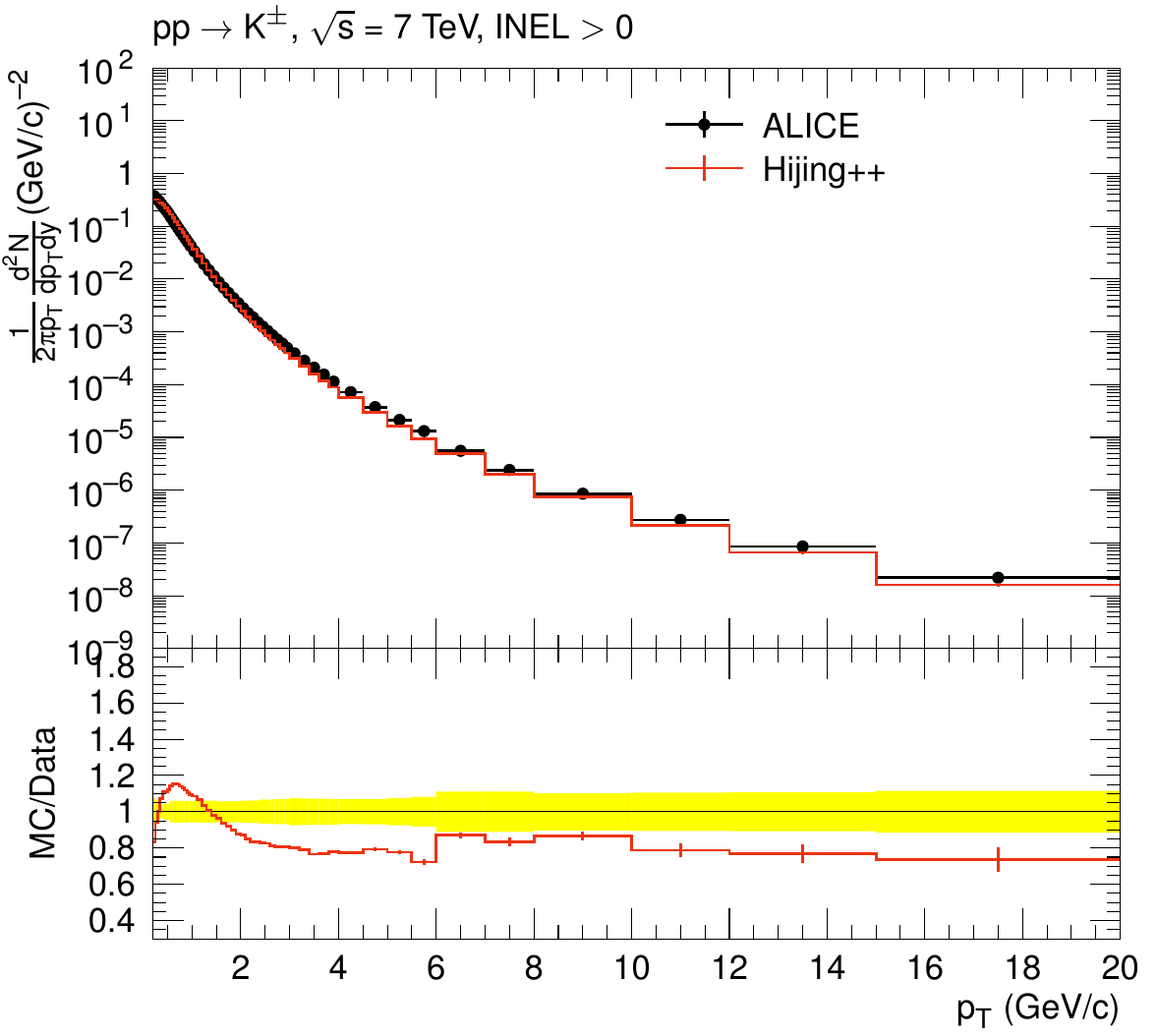}
    \includegraphics[width=0.32\textwidth]{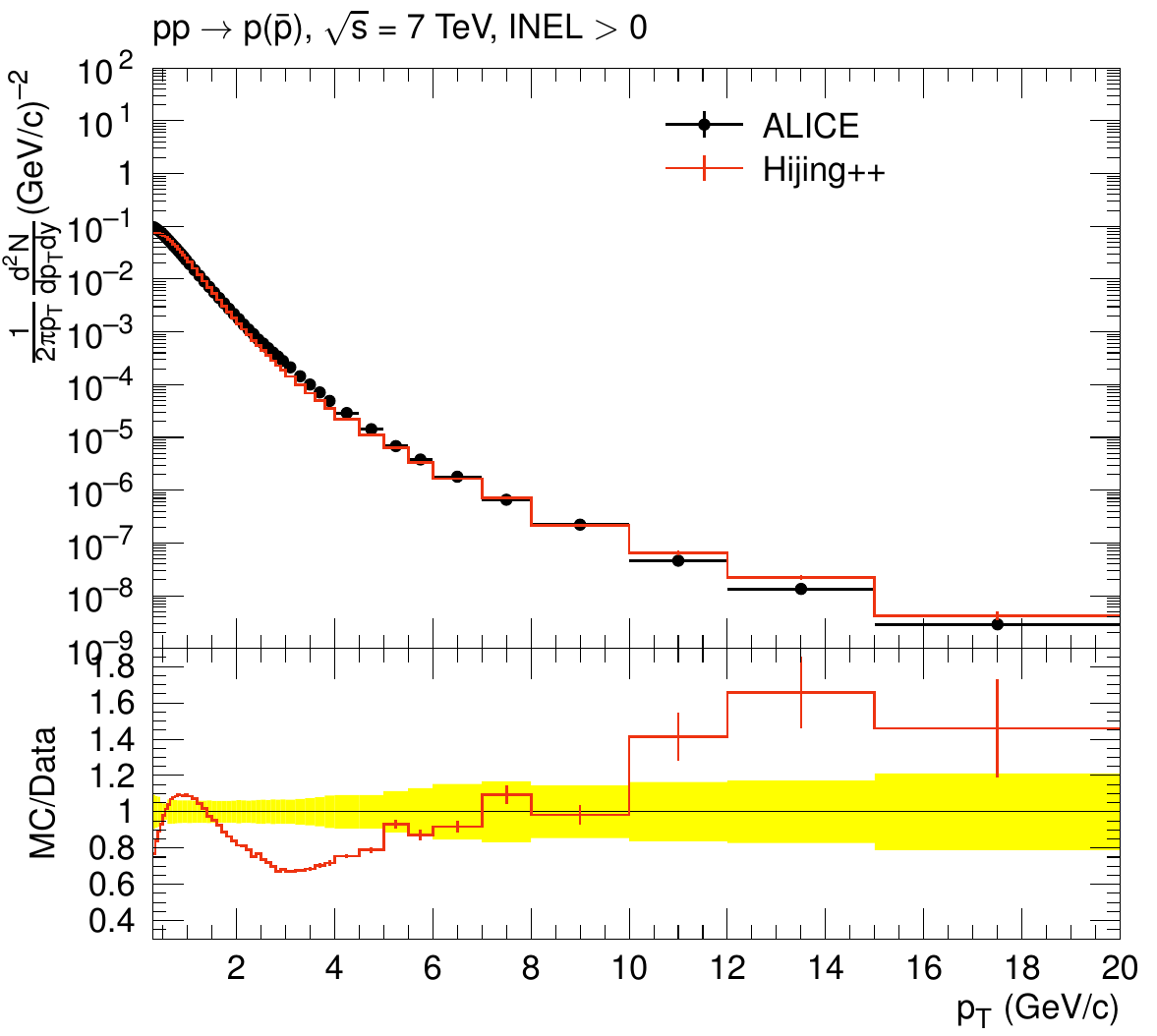}
    \caption{The $p_T$ spectrum of identified $\pi^\pm$ (\textbf{{left panel}}), $K^\pm$ (\textbf{{middle panel}}), and $p(\bar{p})$ (\textbf{{right panel}}) hadrons yield from proton--proton collisions at $\sqrt{s}=7$ TeV with $INEL>0$ normalization calculated with \texttt{HIJING++} and compared to experimental data~\cite{ALICE:PID}.}
    \label{fig:hijing_pid_hadrons}
\end{figure}

In Figure \ref{fig:multiplicity}, the multiplicity and the pseudorapidity distribution of charged hadrons are presented, while in Figure \ref{fig:hijing_pid_hadrons} the $p_T$ spectrum of identified $\pi^\pm$, $K^\pm$, and $p(\bar{p})$ hadrons calculated and measured at $\sqrt{s}= 7$ TeV proton--proton collisions can be seen.

The results above show that \texttt{HIJING++} reproduces the event multiplicity excellently. In Figure \ref{fig:multiplicity}, the agreement between the \texttt{HIJING++} results and the experimental data is $\sim$15\% for the multiplicity and $\sim$1\% for the pseudorapidity distribution. The charged pion and kaon spectra also show a good agreement above $p_T=2$ GeV/c, but the production is slightly overestimated at lower $p_T$ values. The~best agreement for the $\pi^\pm$ results is $\sim$1\% between 2  and 15 GeV/c. For kaons, the yield is slightly underestimated above 2 GeV/c, where the agreement is $\sim$15--20\%. On the other hand, the proton yield is overestimated in the large $p_T$ region, the agreement is $\sim$20--30\%. 


\section{The Non-Extensive Hadronization Model}
\label{sec:nonext}

It is a well-known and an intensively studied phenomenon that the transverse momentum spectra of hadrons stemming from high-energy particle collisions can be described by Tsallis--Pareto type distributions~\cite{BIROG:TSALLIS1, BIROG:TSALLIS2, BIROG:TSALLIS3, BIROG:TSALLIS4, TAKACS:TSALLIS, GRIGORYAN:TSALLIS, ZHU:TSALLIS, KHUNTIA:TSALLIS, WILK:TSALLIS, CLEYMANS:TSALLIS}. Although this observation itself has further consequences, the theory has even more subtle details because of the observed non-trivial dependence on the center-of-mass energy and hadron mass. In the following sections, we show that the parameters also depend on the event multiplicity, i.e., on the size of the system.

We adopted the usual blast-wave assumptions regarding the system, namely that the fireball is azimuthally symmetric and is expanding with a $v$ radial flow velocity {(in units of $c=1$)}. Moreover, the freeze-out occurs instantly on a hypersurface according to the Cooper\,--\,Frye formulation at a given freeze-out temperature~\cite{GRIGORYAN:TSALLIS, URMOSSY:TSALLIS}. With these assumptions, we used the following simple form of the invariant yield:

\begin{equation}
    \left.\frac{d^2 N}{p_Tdp_Tdy}\right|_{INEL>0}=A\cdot m_T\cdot\left(1+\frac{E}{nT} \right)^{-n} \ \ \
    \label{eq:tsallis}
\end{equation}
where $A$ is the amplitude incorporating the irrelevant spin degeneracy and constant factors as well as the invariant volume, $m_T=\sqrt{p_T^2+m^2}$ is the transverse mass, $E=\gamma(m_T-v p_T)-m$ is the one-particle energy in the co-moving coordinate system, $\gamma=1/\sqrt{1-v^2}$ {is the Lorentz factor}, $T$ is a parameter with a temperature unit, and finally $n=\frac{1}{q-1}$ 
is the non-extensivity parameter, characterizing the temperature fluctuations. We note that $T$ is not necessarily the freeze-out temperature and therefore is not necessarily the same for all hadron species~\cite{ZHU:TSALLIS, VAN:TSALLIS}. The notation $INEL>0$ means that only those events  where there is at least one charged particle in the $|\eta|<1.0$ region are considered. This choice is in agreement with the experimental definitions described in the previous section.

\begin{figure}[H]
    \centering
    \includegraphics[width=0.329\textwidth]{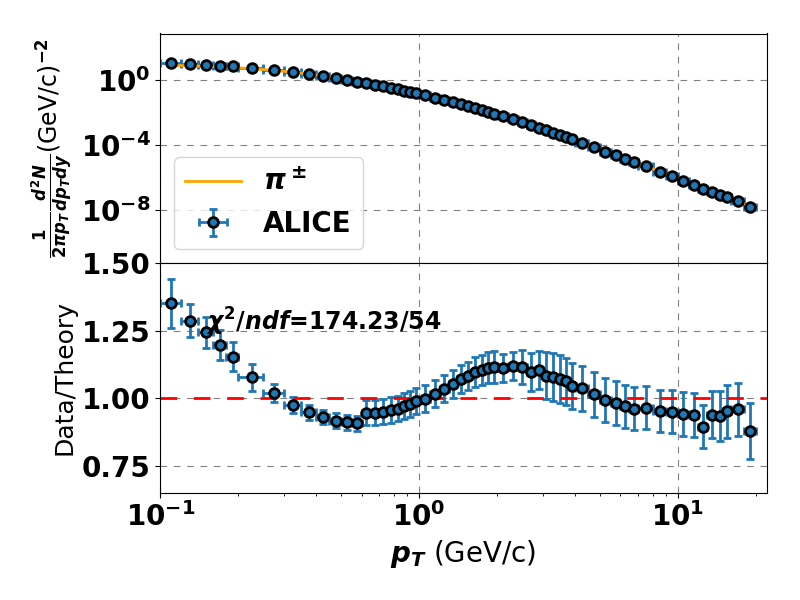}
    \includegraphics[width=0.329\textwidth]{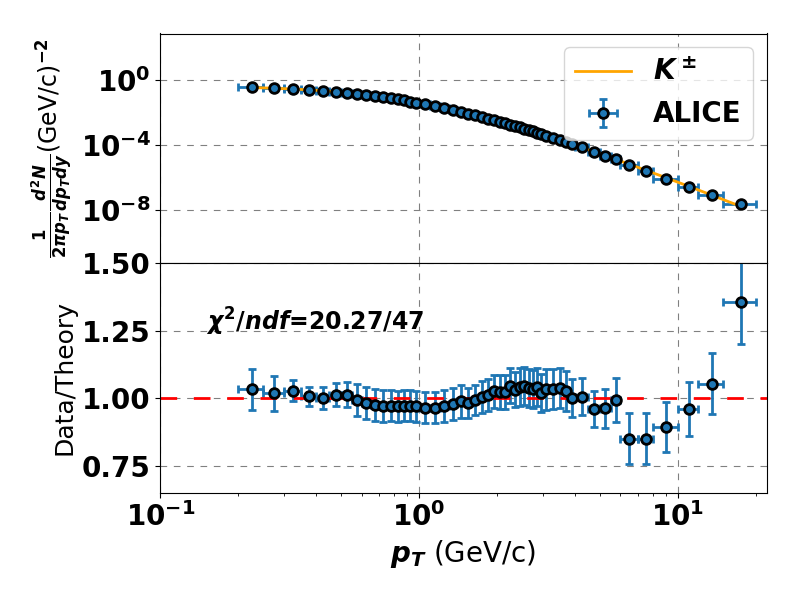}
    \includegraphics[width=0.329\textwidth]{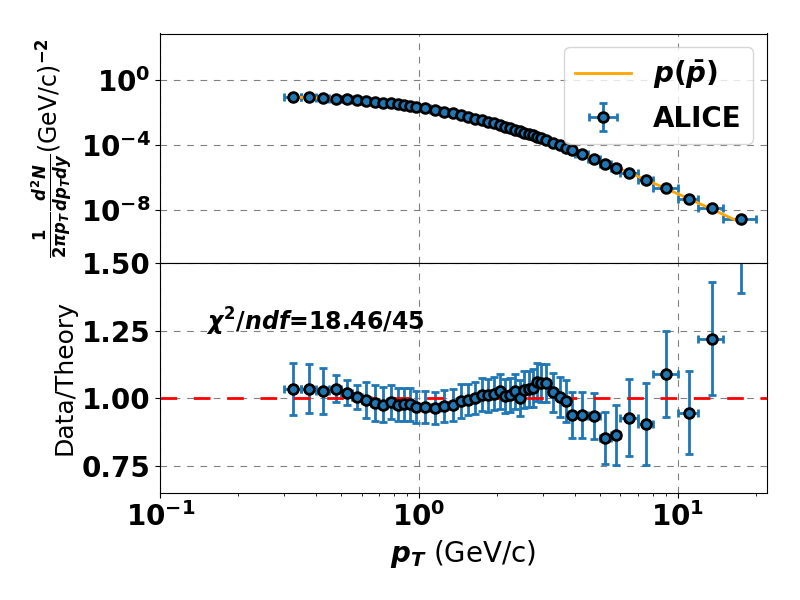}
    \caption{Fits of the Tsallis--Pareto distribution to $\pi^\pm$, $K^\pm$, and $p(\bar{p})$ hadrons measured by ALICE~\cite{ALICE:PID}.}
    \label{fig:exp_fits}
\end{figure}

As a reference, in Table \ref{tab:alice_parameters} and in Figure \ref{fig:exp_fits}, we show the parameters and curves fitted on the experimental ``minimum bias'' (in the sense that there is no event multiplicity classification) data. These results are consistent with our previous observations~\cite{BIROG:TSALLIS1}: the heaviest proton has the largest temperature and the smallest $q$. {We note that, in the lower part of Figure} \ref{fig:exp_fits}, {a periodic oscillation is visible. This is an effect in addition to the scaling. This has been  
 investigated, for example, in   }\cite{WILK:OSCILLATIONS, RYBCZYNSKI:OSCILLATIONS}.

\begin{table}[H]
    \caption{Tsallis parameters extracted from ``{minimum bias}'' $INEL>0$ proton--proton {collisions at} $\sqrt{s}=7$ TeV, measured by ALICE~\cite{ALICE:PID}}
    \label{tab:alice_parameters}
    \centering
    \begin{tabular}{ccccccc}
        \toprule
        \textbf{Hadron} &  \boldmath{$n$} & \boldmath{$q$} & \boldmath{$T$} \textbf{(GeV)} & \boldmath{$A$} & \boldmath{$v$} & \boldmath{$\chi^2/ndf$} \\
        \midrule
        \textsc{ $\pi^\pm$ } &  7.415 $\pm$ 0.033  &  1.135 $\pm$ 0.005  &  0.089 $\pm$ 0.010  &  73.188 $\pm$ 9.700  &  0.000 $\pm$ 0.119  &  174.225 / 54 \\
        \textsc{ $K^\pm$ } &  7.539 $\pm$ 0.086  &  1.133 $\pm$ 0.013  &  0.155 $\pm$ 0.010  &  0.915 $\pm$ 0.095  &  0.000 $\pm$ 0.066  &  20.274 / 47 \\
        \textsc{ $p(\bar{p})$ } &  8.805 $\pm$ 0.184  &  1.114 $\pm$ 0.023  &  0.191 $\pm$ 0.012  &  0.124 $\pm$ 0.013  &  0.000 $\pm$ 0.054  &  18.462 / 45 \\
        \bottomrule
\end{tabular}
\end{table}



\section{The Multiplicity Dependence of the Non-Extensive Model}
\label{sec:multiplicity}

In Section \ref{sec:tuning}, we showed that the \texttt{HIJING++} framework is  able to reproduce the main experimental observables such as event multiplicity distribution and the $p_T$ spectra of various identified hadrons. In Section \ref{sec:nonext}, we briefly summarized the main features of the blast-wave motivated non-extensive hadronization model. In this section, we take advantage of the power of \texttt{HIJING++} and extract the Tsallis parameters from a wide range of event multiplicity classes. The event classes of the \texttt{HIJING++} run are classified as 

\begin{equation}
\textrm{Class}=\left<dN_{ch}/d\eta\right>_{min}<\left<dN_{ch}/d\eta\right>\leq\left<dN_{ch}/d\eta\right>_{max} .
\label{eq:mclass}
\end{equation}

The multiplicity ranges of each class used in this study are listed in Table~\ref{tab:mult_ranges}.
\begin{table}[H]
    \caption{Multiplicity classes used in HIJING++ runs.}
    \label{tab:mult_ranges}
    \centering
    \begin{tabular}{cccccccccccccccc}
        \toprule
        \textbf{Class}	& \textsc{\textbf{I}} & \textsc{\textbf{II}} &\textsc{\textbf{III}} &\textsc{\textbf{IV}} &\textsc{\textbf{V}} &\textsc{\textbf{VI}} &\textsc{\textbf{VII}} &\textsc{\textbf{VIII}} &\textsc{\textbf{IX}} &\textsc{\textbf{X}} &\textsc{\textbf{XI}} &\textsc{\textbf{XII}} &\textsc{\textbf{XIII}} &\textsc{\textbf{XIV}} &\textsc{\textbf{XV}}	\\
        \midrule
        \textbf{$\left<dN_{ch}/d\eta\right>_{min}$} & 0& 10& 15& 20& 25& 30& 35& 40& 45& 50& 55& 60& 70& 80& 90 \\
        \textbf{$\left<dN_{ch}/d\eta\right>_{max}$} & 10 & 15 & 20 & 25 & 30 & 35 & 40 & 45 & 50 & 55 & 60 & 70 & 80 & 90 & 100\\
        \bottomrule
    \end{tabular}
\end{table}

\begin{figure}[H]
    \centering
    \includegraphics[width=0.325\textwidth]{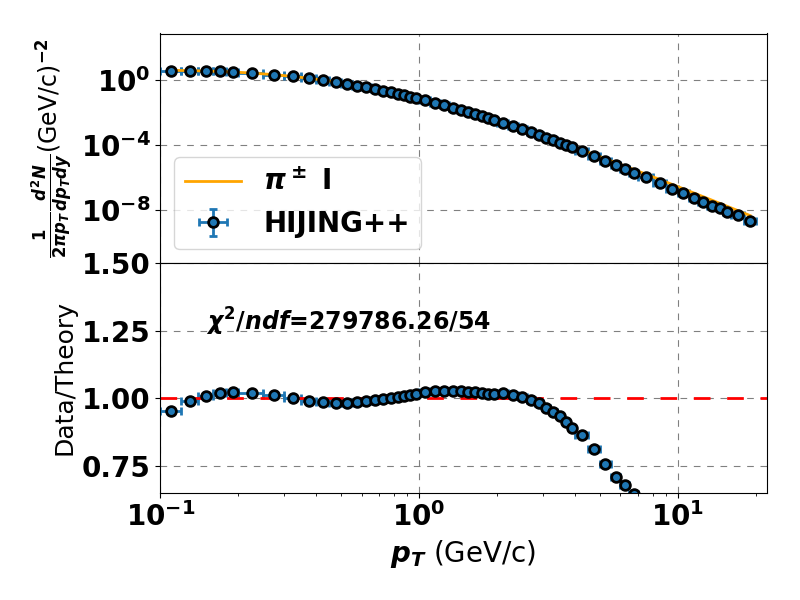}
    \includegraphics[width=0.325\textwidth]{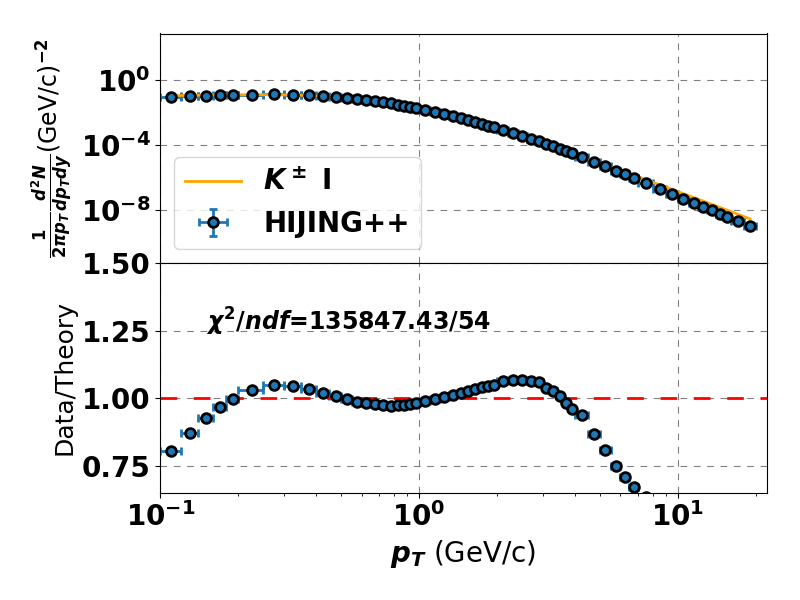}
    \includegraphics[width=0.325\textwidth]{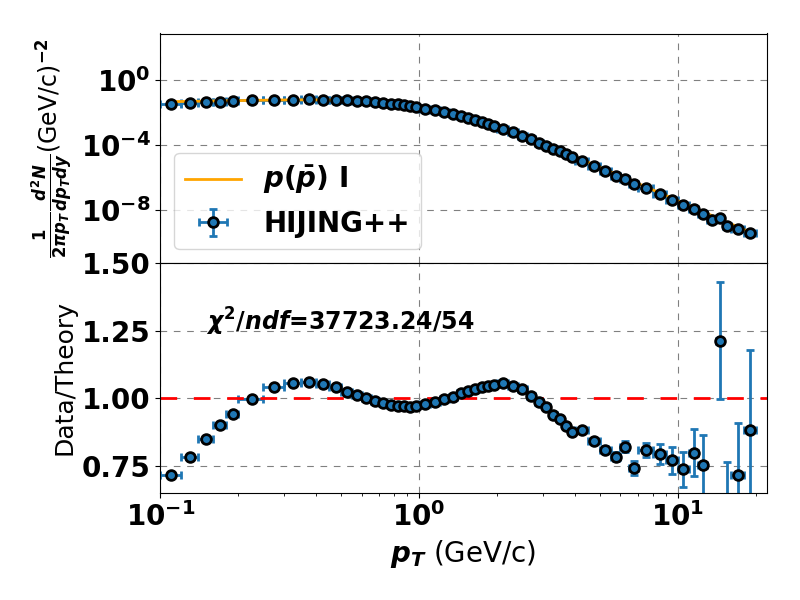}

    \includegraphics[width=0.325\textwidth]{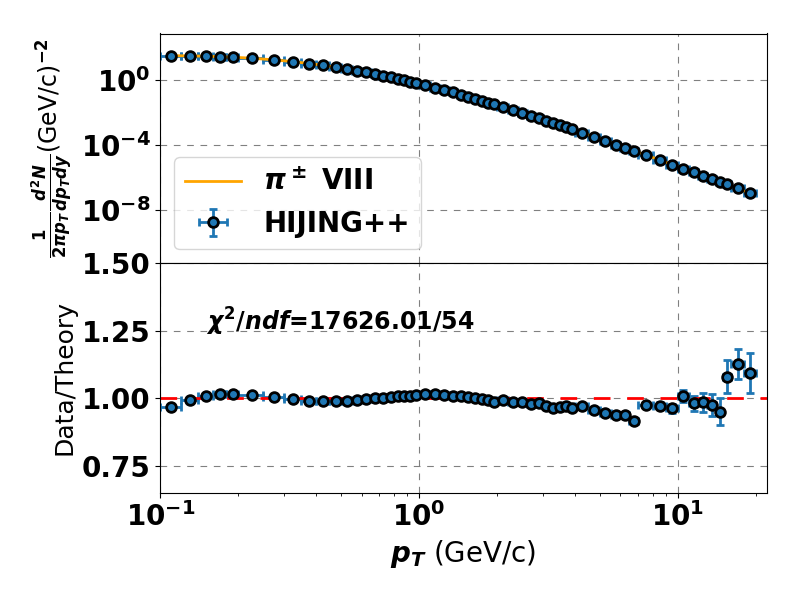}
    \includegraphics[width=0.325\textwidth]{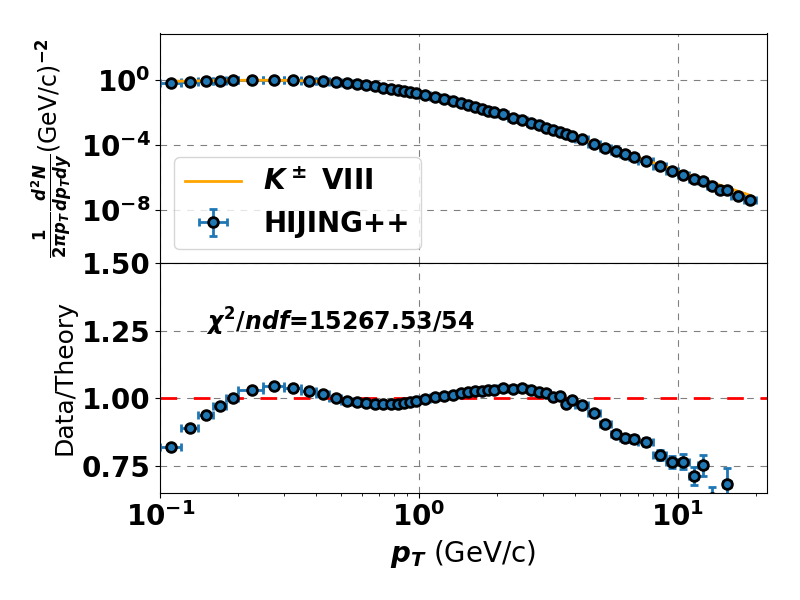}
    \includegraphics[width=0.325\textwidth]{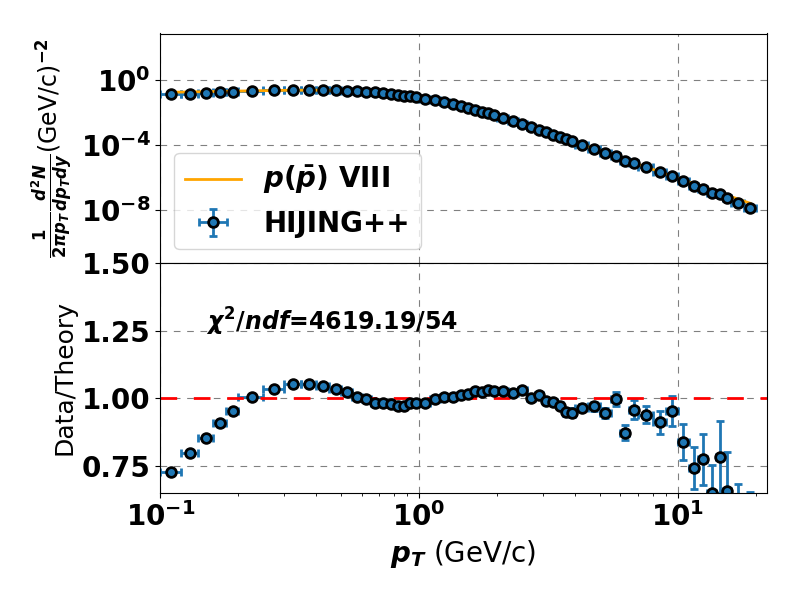}

    \includegraphics[width=0.325\textwidth]{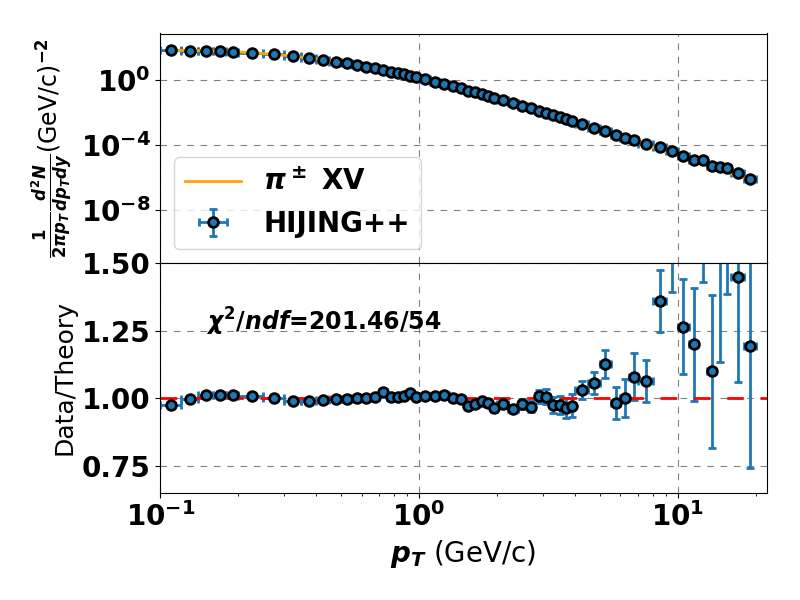}
    \includegraphics[width=0.325\textwidth]{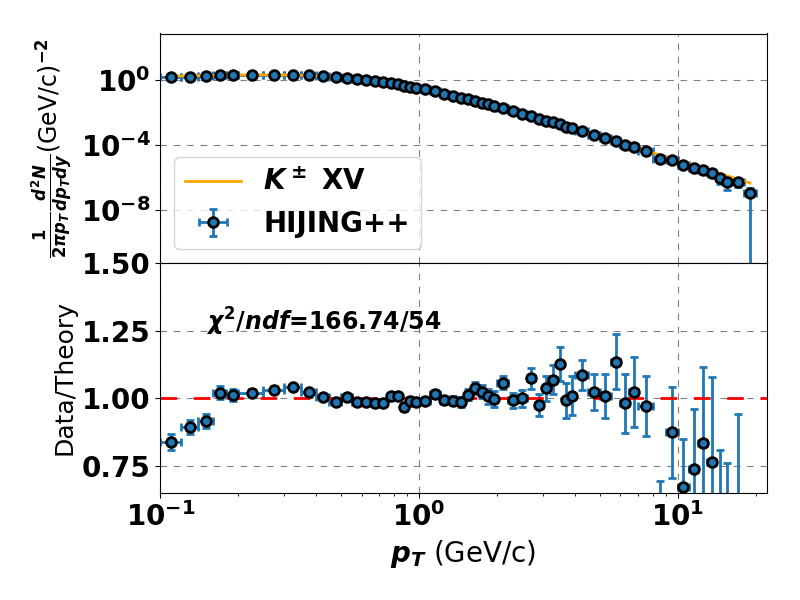}
    \includegraphics[width=0.325\textwidth]{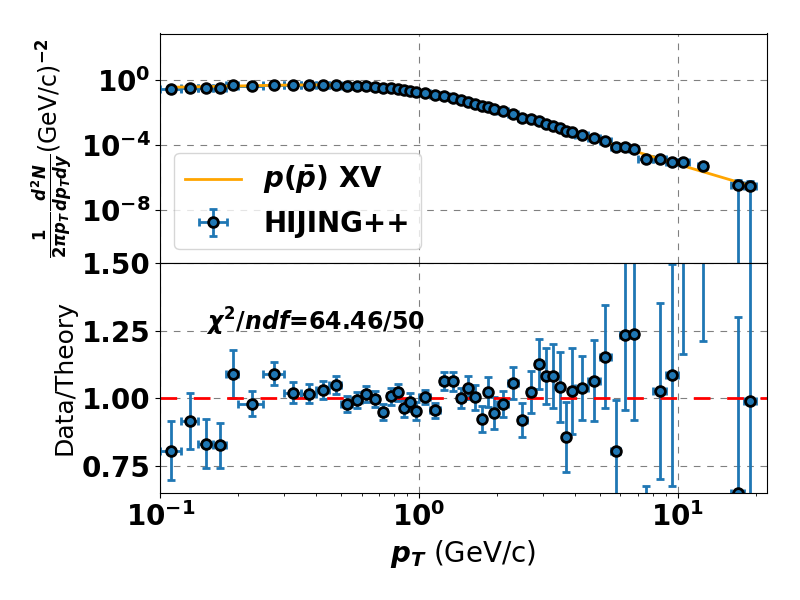}
    \caption{Calculated $p_T$ spectra of charge averaged pions (\textbf{left column}), kaons (\textbf{middle column}), and~protons (\textbf{right column}) at low (\textbf{top row}), moderate (\textbf{middle row}), and high (\textbf{bottom row}) multiplicity classes as blue dots and the fitted {Equation }(\ref{eq:tsallis}) Tsallis--Pareto curve (orange line). The~lower part of each panel shows the \textit{Data/Theory} ratio.}
    \label{fig:hpp_fits}
\end{figure}

Using this event classification, we calculated the mid-rapidity transverse momentum spectra of charge averaged pions, kaons, and~protons in $INEL>0$ events. We {generated} $200$M events. To~avoid superfluous overcrowding of the available space, we show only the low, moderate, and high multiplicity spectra along with the fitted Tsallis--Pareto curves {defined by Equation }(\ref{eq:tsallis}) in Figure \ref{fig:hpp_fits}. 

In Figure \ref{fig:hpp_fits}, we can see that the best fits occurred at the high multiplicity events. For the pions, the~fitted curves follow the points well at the low-$p_T$ region, while for the kaons and protons with higher mass at the low-$p_T$ region the model overpredicts the yield. We note, however, that for the \texttt{HIJING++} run we investigated the same $0.1$ GeV/c$<p_T<20$ GeV/c region for all hadrons, while the low-$p_T$ part for the kaons and protons in the case of the experimental results is missing. At the high-$p_T$ region, the fit breaks down because of the {low} statistics.

\begin{figure}[H]
    \centering
    \includegraphics[width=0.40\textwidth]{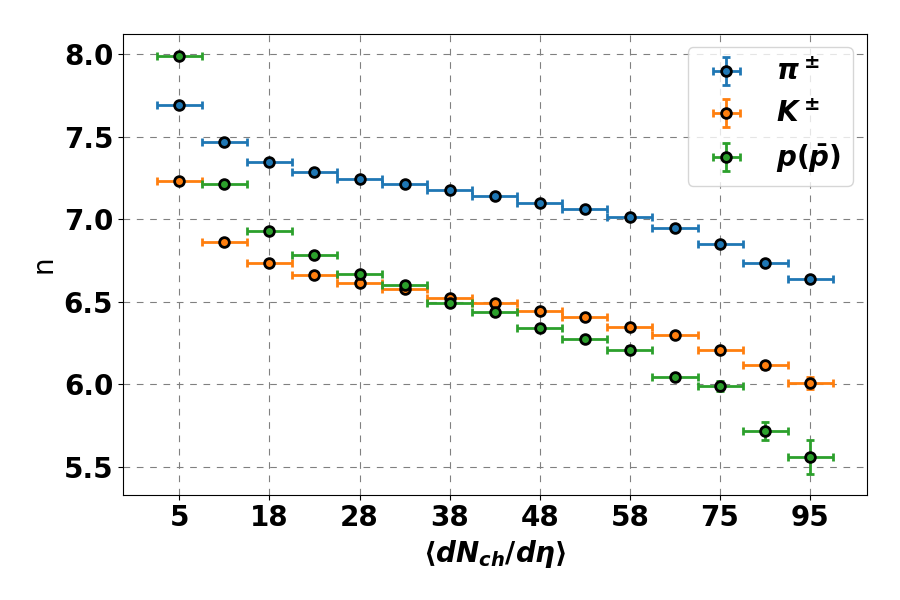}
    \includegraphics[width=0.40\textwidth]{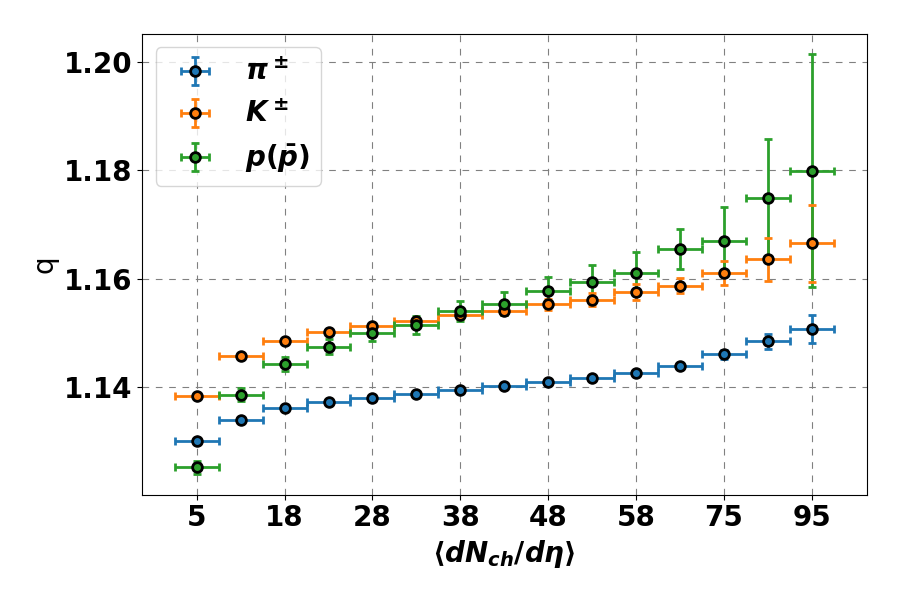}
    \includegraphics[width=0.40\textwidth]{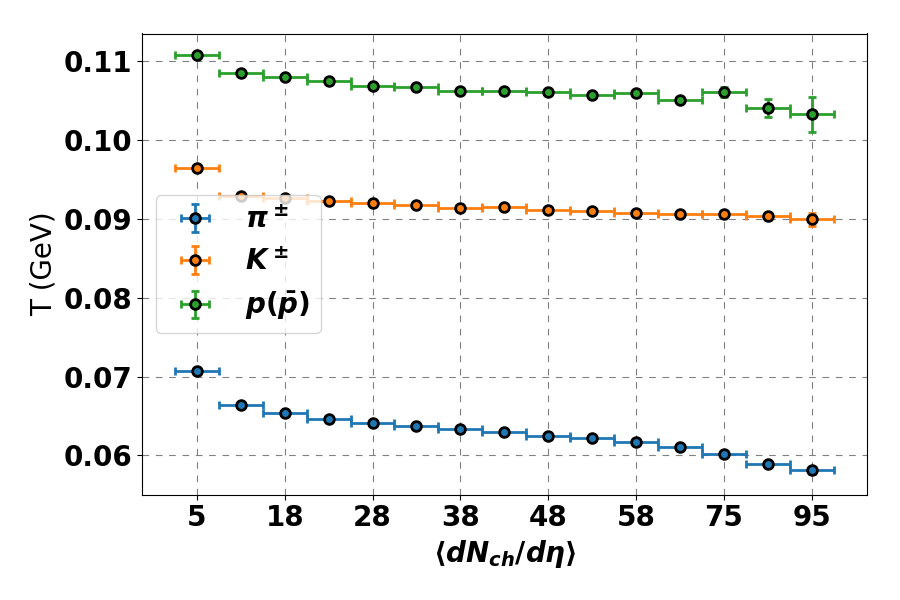}
    \includegraphics[width=0.40\textwidth]{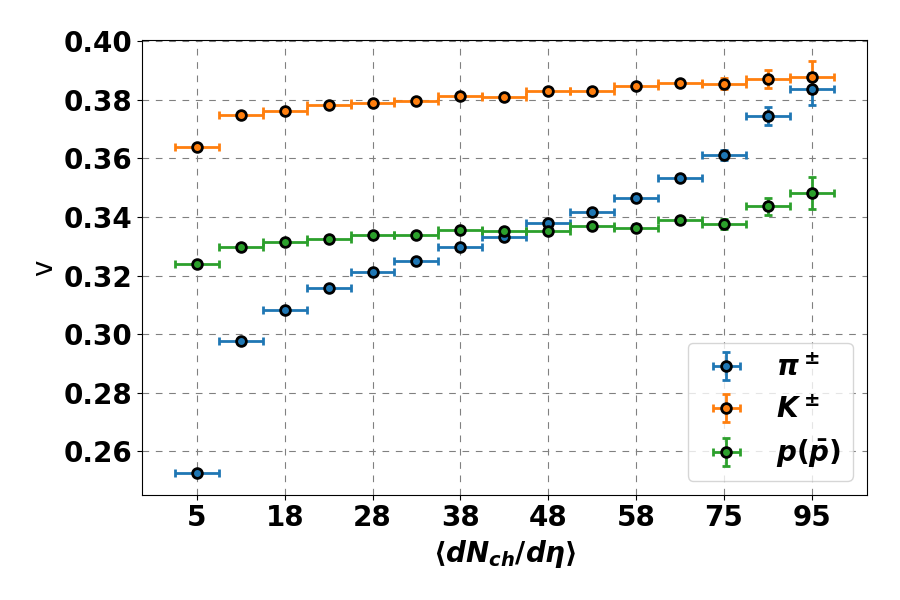}
    \includegraphics[width=0.40\textwidth]{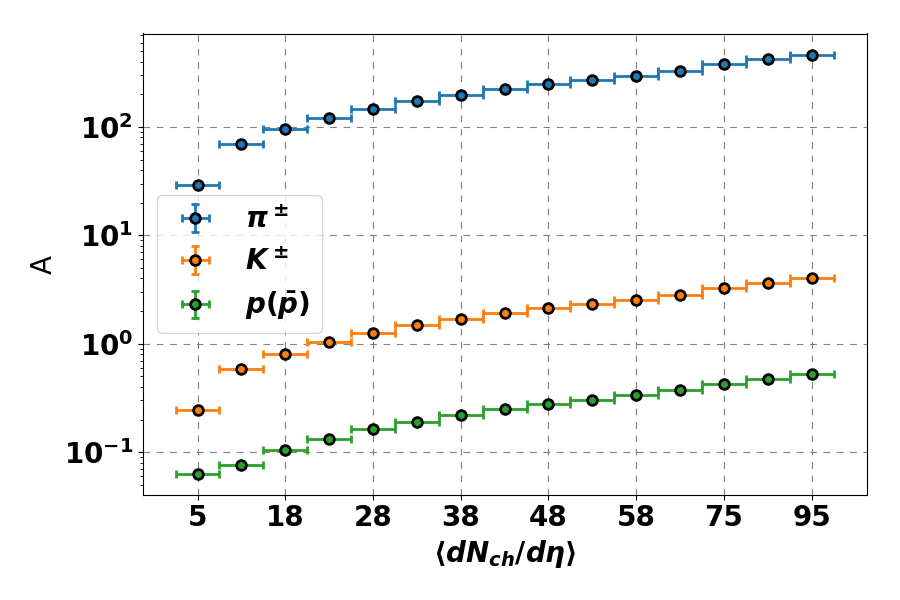}
    \caption{The fitted parameters of the Tsallis--Pareto distribution defined by Equation (\ref{eq:tsallis}), in the function of the event multiplicity class defined as in Table \ref{tab:mult_ranges}.}
    \label{fig:fit_params}
\end{figure}

In Figure \ref{fig:fit_params}, the fitted parameters in the function of the event multiplicity are shown. Using~the distribution form Equation (\ref{eq:tsallis}), we observe that with increasing multiplicity the $q$ parameter increases for each hadron but with different slopes: the increase of $q$ (or the decrease of $n$) is the largest for the heaviest hadron. On the other hand, the temperature decreases slowly with the increasing {pseudorapidity density}. Here, the previously observed $T_{\pi^\pm} < T_{K^\pm} < T_{p(\bar{p})}$ mass hierarchy {stays valid} with the multiplicity averaged values, as   can be seen in Table \ref{tab:avg_temp}.

\begin{table}[H]
  \caption{\textls[-15]{The multiplicity averaged temperature parameters for charge averaged pions, kaons, and protons.}}
  \label{tab:avg_temp}
  \centering
  \begin{tabular}{cc}
      \toprule
      \textbf{Hadron} &  \boldmath{$\left<T_{i}\right>$} \textbf{(GeV)}\\
      \midrule
      $\pi^\pm$  &    $0.063 \pm 0.003$ \\
      $K^\pm$  &      $0.092 \pm 0.001$ \\
      $p(\bar{p})$  & $0.106 \pm 0.002$ \\
      \bottomrule
\end{tabular}
\end{table}

The radial flow velocity also increases with the multiplicity, but also with different rates. While~at low multiplicity the lightest pions have the smallest $v$, it increases rapidly with the increasing multiplicity. On the other hand, the rate of increase in the case of protons and kaons are approximately the same. These observations require further investigation.
Finally, the amplitudes are increasing for each hadron species with the multiplicity. The value of pions is much higher than those of 
 the heavier hadrons, which indicates that with increasing multiplicity the number of the produced pions grows faster than the number of kaons and protons.

\section{Conclusions}

In this contribution, we investigated the multiplicity dependence of the parameters of the non-extensive hadronization model in proton--proton collisions using \texttt{HIJING++} calculations. We~presented the current status of the tuning process of \texttt{HIJING++} and showed that it is able to reproduce the main high-energy physics observables such as multiplicity and $p_T$ distributions. We~presented the non-extensive hadronization model that we used to describe the transverse momentum distribution of identified hadrons. In accordance with our previous results, we showed that a mass hierarchy emerges in the Tsallis parameters. Utilizing the tuned \texttt{HIJING++} calculations, we also extracted the parameters   from $\sqrt{s}=7$ TeV proton--proton {collisions} Monte Carlo calculations with various event multiplicity classifications. Our study showed that the $q$ non-extensivity parameter increases with   increasing multiplicity, while the $T$ temperature has only a slight decrease. On the other hand, all~hadrons result in a non-zero, increasing radial flow velocity. All parameters show   the earlier observed mass hierarchy. Our findings suggest that these parameters are sensitive to the event size and may serve as a thermometer of the collision. 

\vspace{6pt} 



\authorcontributions{Software and formal analysis: G.B. and G.P.; investigation: G.B. and G.G.B.; writing---original draft preparation: G.B.; writing---review and editing: G.G.B.; supervision: G.G.B., T.S.B., and G.P.; funding acquisition: G.G.B. and T.S.B.
  }

\funding{This research was funded by Hungarian-Chinese cooperation grant No. MOST 2014DFG02050, the~Wigner HAS-OBOR-CCNU grant, and OTKA grants K120660 and K123815
. G.B. acknowledges the support of  the Wigner Data Center and Wigner GPU Laboratory.}


\conflictsofinterest{The authors declare no conflict of interest.} 

\reftitle{References}
\bibliographystyle{unsrtnat}





\end{document}